# Fast and accurate detection of Covid-19-related pneumonia from chest X-ray images with novel deep learning model

M. M. Ramadhan[1,8], A. Faza[1,8], L. E. Lubis[1,a], R. E. Yunus[2,3], T. Salamah[2,3], D. Handayani[4,5], I. Lestariningsih[6], A. Resa[6], C. R. Alam[6], P. Prajitno[1], S. A. Pawiro[1], P. Sidipratomo[2,7], and D. S. Soejoko[1, a)]

[1.] *Department of Physics, Faculty of Mathematics and Natural Sciences, Universitas Indonesia, Depok 16424, Indonesia*
[2.] *Department of Radiology, Faculty of Medicine, Universitas Indonesia, Jakarta 10430, Indonesia*
[3.] *Radiology Unit, Universitas Indonesia Hospital, Depok, 16424, Indonesia*
[4.] *Department of Pulmonology and Respiratory Medicine, Faculty of Medicine, Universitas Indonesia, Jakarta 10430, Indonesia*
[5.] *Pulmonology Unit, Universitas Indonesia Hospital, Depok, 16424, Indonesia*
[6.] *Department of Radiology, Cibinong Regional General Hospital, Bogor, 16914, Indonesia*
[7.] *Faculty of Medicine, UPN Veteran, Jakarta 12450, Indonesia*
[8.] *Noobtech Indonesia, Depok, 16424, Indonesia*

a) Corresponding author: lukmanda.evan@sci.ui.ac.id

**Background:** Novel coronavirus disease has spread rapidly worldwide. As recent radiological literatures on Covid-19 related pneumonia is primarily focused on CT findings, the American College of Radiology (ACR) recommends using portable chest X-radiograph (CXR). A tool to assist for detection and monitoring of Covid-19 cases from CXR is highly required.

**Purpose:** To develop a fully automatic framework to detect Covid-19 related pneumonia using CXR images and evaluate its performance.

**Materials and Methods:** In this study, a novel deep learning model, named CovIDNet (Covid-19 Indonesia Neural-Network), was developed to extract visual features from chest x-ray images for the detection of Covid-19 related pneumonia. The model was trained and validated by chest x-rays datasets collected from several open source provided by GitHub and Kaggle.

**Results and Discussion:** In the validation stage using open-source data, the accuracy to recognize Covid-19 and others classes reaches 98.44%, that is, 100% Covid-19 precision and 97% others precision.

**Discussion:** The use of the model to classify Covid-19 and other pathologies might slightly decrease the accuracy. Although SoftMax was used to handle classification bias, this indicates the benefit of additional training upon the introduction of new set of data.

**Conclusion:** The model has been tested and get 98.4% accuracy for open source datasets, the sensitivity and specificity are 100% and 96.97%, respectively.

## A. INTRODUCTION

The Covid-19 outbreak is an unprecedented global public health challenge. At least, until March 16 2020, there were 210 affected countries, including Indonesia. In Indonesia, the first cases were recorded at March 1 and at April 16 the number of cases had soared to 5.923, with mortality rate reaching an alarming number of 8.78%. A technology to detect Covid-19 quickly and accurately is required with utmost urgency. Currently, this novel disease is confirmed by reverse polymerase transcription chain reaction (RT-PCR) tests. However, it has been reported that RT-PCR sensitivity may not be high enough for this purpose early detection and may need complementary tests. Computed tomography (CT), as a non-invasive imaging approach, is able to detect manifestations of certain characteristics in the lung associated with Covid-19. Owing to that reason, Covid-19 radiological literatures during the early outbreak in Wuhan, China, were primarily focused on CT findings [1]. Recently, artificial intelligence (AI) using deep learning has shown great success in the medical imaging domain because of its high feature extraction ability. It is deemed useful for many tasks, namely classification, denoising, prediction, and auto detection. In the beginning of 2020, several researchers from Wuhan, China proposed methods to distinguish Covid-19 with a new deep learning model namely, COVNet (Covid-19 Detection Neural Network) using chest CT images. The model divides into three class, namely Covid-19, Community-Acquired Pneumonia (CAP), and other non-Pneumonia cases. The sensitivity and specificity reached 90% and 96% respectively [2].

However, the use of CT to detect Covid-19 as a first-line diagnosis presents a massive burden on radiology departments and posed an immense challenge for infection control in the CT suite. The American College of Radiology (ACR) notes that CT decontaminations are required after scanning of Covid-19 patients and it will interfere with other radiological services while other patients are prone to hospital-acquired infections of Covid-19. Thus, ACR does not recommend CT as first-line screening. Furthermore, ACR recommends using portable chest radiograph (CXR), as it is believed to minimize the risk of cross infection owing to possible device dedication due to high availability of mobile x-ray devices.

The previous work to detect Covid-19 by CXR images was performed by L. Wang and A. Wong from Canada, in which they build new deep learning model named Covid-Net. They divide the model into 3 class, namely Covid-19, normal, and pneumonia with accuracy reaching 88.2%. They subsequently publish their dataset on GitHub and Kaggle [3], [4]. At the open source dataset, it used same dataset as Covid-Net model. The aim of this study is to provide fast detection tool of Covid-19 by CXR images with deep learning model with high accuracy and lowest false negative as paramount interest.

## B. METHOD

### B.1. Preprocessing

Every image from different machine and center has different parameters, i.e. hue, saturation, contrast, and size. Preprocessing is, therefore, necessary to normalize the images into a uniform format. Images were collected as JPEG, JPG, PNG, and DICOM formats. At the first preprocessing step, images were cropped to clear borders and make sure that the images only include lungs. Afterwards, contrast-limited adaptive histogram equalization (CLAHE) method was applied to reduce noise amplification. In CLAHE method, the contrast amplification in the vicinity of a given pixel value is given by the slope of the transformation function. At the end of preprocessing, by still considering the aspect ratio of the images, all images were resized to the same size. For images in DICOM format, it applied rescale (Modality LUT) and windowing operation (VOI LUT) prior to preprocessing.

## B.2. Deep Learning Model

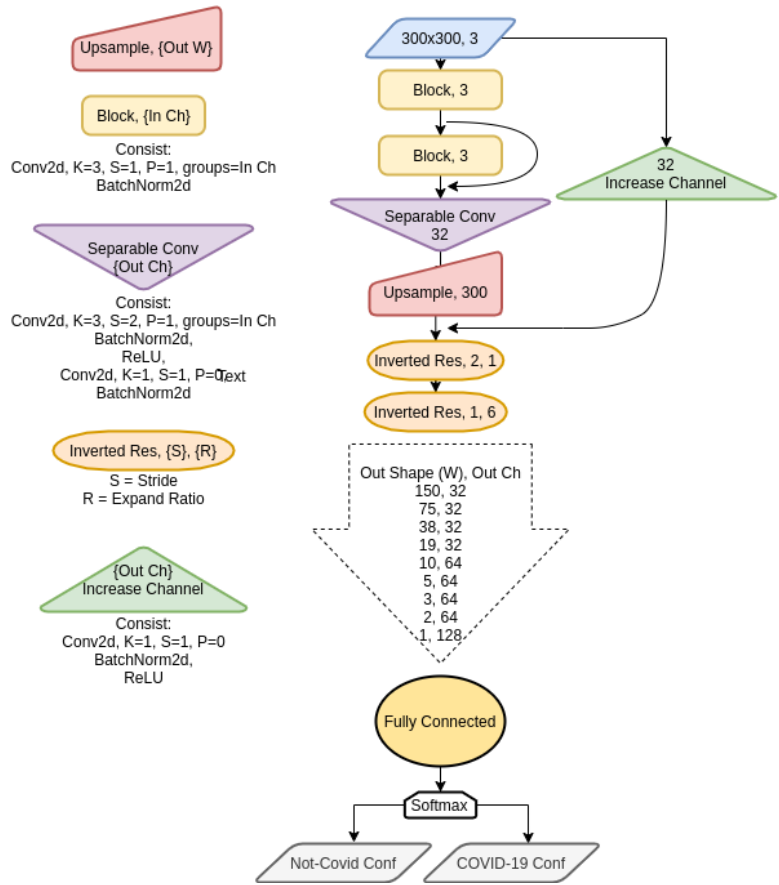

*Figure 1. CovIDNet Architecture Deep Learning Model*

Our deep learning architecture (Figure 1), named CovIDNet, was built to extract feature of chest x-ray image and detect Covid-19 related pneumonia from suspects. Our model is adapted from MobileNetV2 and ResNext, that has been widely used for many purposes in object classification and detection, therefore serving as a reasonable method. As shown on Figure 2, our model utilizes Inverted Residual and Separable Convolution which are state-of-the-art convolution from MobileNetV2. MobileNetV2 had shown its performance with less computation and competitive results [5]. Our model also used skip connection with summation fashion that was inspired by ResNext. ResNext performance has overcome his ancestor ResNet [6] and also was the First Runner up in ILSVRC 2016 [7]. ResNext shown skip connection with summation yield in better result that overcome another skip connection technique with concatenate fashion (i.e. Inception from ResNet) [8]. Inspired by some method of those previous works, CovIDNet consist of 580015 parameters that was conducted on light amount of memory to forward the model compared to. Resnext and MobilenetV2 who have 22345027 and 4038499 parameters respectively. Referring to our original purpose, CovIDNet is aimed to act with less parameters to avoid overfitting and high computational load requirement.

From the output of CovIDNet, SoftMax was utilized to make decisions, i.e. with confidence score (range 0-1) and provide prediction. To determine the validity of the prediction, the model have criterion. This criterion is being the value of confidence score, with confidence score below 0.5 being considered as invalid. This approach was taken to avoid false negative result and denotes users to take other diagnostic tools.

### B.3. Training

The model was trained with two class models. However, requiring three output classes was proven to be a challenge. Previous works have done classification by adding one more class (called 'background') that did not refer to any class of the trainset [9]. A reason for adding this background class was to aid the initial training process and as solution to imbalance dataset. At first epoch, the model has no prior knowledge in deciding to which category the data falls into. Consequently, the model prediction confidence level spread randomly into each class that was accounted to high loss on initial process. The additional output would gather some confidence and minimize loss on another class.

Hence, training will focus only on output that has data class to train for. In addition, it also performed hard negative mining, i.e. a filter to take negative to positive with some ratio. With this mechanism the number of negative predictions could be suppressed [10]. Imbalanced trainset has problem with large number of negative predictions since the model tends to fit to the dominant class and hard negative mining is a solution to this problem that used on *Single Shot MultiBox Detector* (SSD) training [11]. In the application, the additional class would not be accounted in prediction since after training process the model would not predict with high confidence on this class.

The training and validation were divided into two steps based on the datasets, namely primary and advance step. At the primary step, the dataset used has also been used in previous works to formulate the Covid-Net model. It is followed by open source data from two links, namely: *https://github.com/ieee8023/covid-chestxray-dataset* and *https://www.kaggle.com/c/rsna-pneumonia-detection-challenge*. The class were divided into two, namely 'Covid19' and 'others', with 'others' being a set of CXR that has been diagnosed as normal and another lung pathology. The ratio of training and testing in terms of number of images was 75:25. Moreover, image augmentation was done to multiply the data, while the augmentation technique used are random horizontal flip, random CLAHE, and random color jitter. Limiting technique of augmentation is a must to consider the state of the image in the real field.

## C. RESULTS

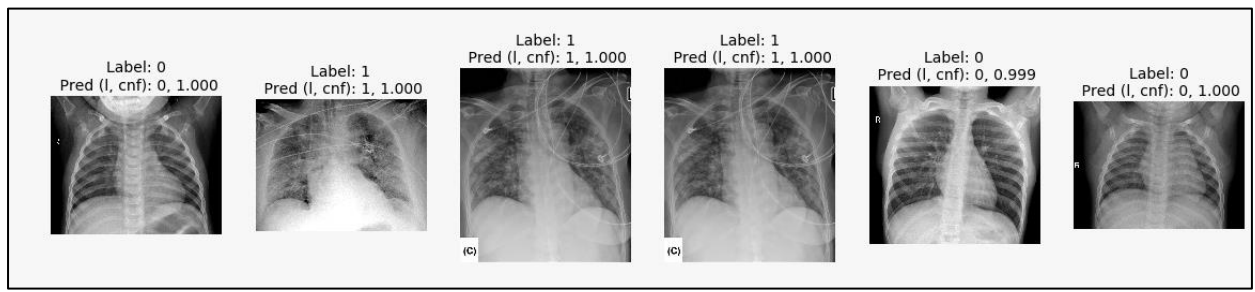

*Figure 2. Classification of six random CXR images*

Figure 3 displays the results of the evaluation step by taking six random chest x-ray images into each class. It calculated the level of accuracy based on the confusion matrix, i.e. a table used to describe the performance of a classification model (or "classifier") on a test data set whose true value is known. This allows visualization of the performance of an algorithm. In the primary stage, the accuracy to recognize Covid-19 and others classes reaches 98.44%, that is, 100% Covid-19 precision and 97% others precision. Furthermore, the specificity is 96.97% and the sensitivity is 100%. This results are currently higher than previous studies to classify Covid-19 based on chest x-ray images [3][12]. when the model is used to recognize 'Others' and 'Covid-19', difficulty is present with only two model classes while one being of many types of pathology with similar physical characteristics in the lungs with Covid-19. To handle bias classification and false negative, SoftMax was used. However, different x-ray machines might have different x-ray beam quality and imaging characteristics due to different type of image receptors. Therefore, additional data training might be beneficial prior to applying this model on new set of devices. To facilitate further use, this model is run in web form with the link *http://sci.ui.ac.id/detectcovid* with the interface being shown in Figure 3, the web can diagnosis Covid-19 by this novel deep learning model and only require less than 1 minute processing time.

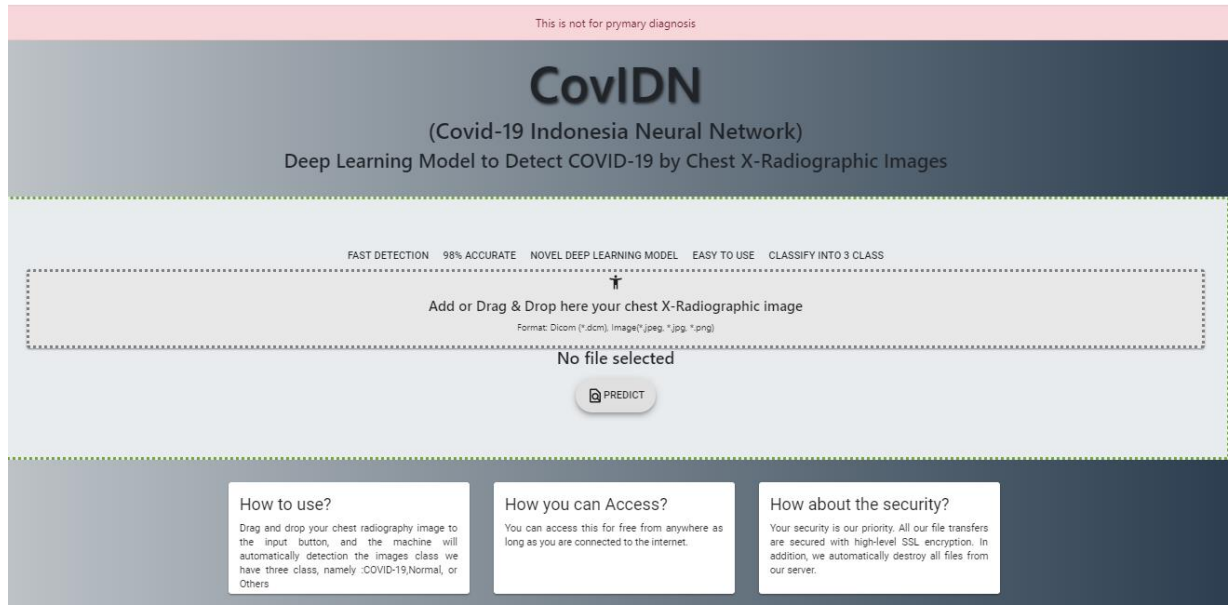

*Figure 3. Interface of web-based Diagnosis Support System (DSS) CovIDNet in http://sci.ui.ac.id/detectcovid*

## D. CONCLUSION

A novel deep learning model has been developed to detect Covid-19-related pneumonia cases and differentiate it from normal CXR and other lung pathologies. The model has been tested with 98.44% accuracy for open source datasets, the sensitivity and specificity are 100% and 96.97%, respectively.

## ACKNOWLEDGMENTS

This study is collaborative research between Artificial Intelligence for Radiological Application (AIRA) research group in Department of Physics, Faculty of Mathematics and Natural Sciences, Universitas Indonesia, Universitas Indonesia Hospital, Cibinong Regional General Hospital, and Noobtech Indonesia. The authors declare no conflict of interest and financial implications related with this work.